\documentclass[aps,pra,twocolumn,halfspacing,10pt,nofootinbib]{revtex4-2}
\usepackage{epsfig}
\usepackage{dcolumn}
\usepackage{natbib}
\usepackage[colorlinks]{hyperref}
\usepackage[utf8]{inputenc}
\usepackage[english]{babel}
\usepackage{graphicx}
\usepackage{color}
\usepackage{bm}
\usepackage{amssymb}
\usepackage[intlimits]{amsmath}
\usepackage{booktabs}
\usepackage{amsmath}
\usepackage[para,online,flushleft]{threeparttable}

\usepackage{physics}

\begin{document}
\title{Enhanced  Magnetic Quadrupole Moments in Nuclei with Octupole Deformation and their CP-violating effects in molecules }
\author{V. V. Flambaum$^{1,2}$} 
\email{v.flambaum@unsw.edu.au}
\author{A. J. Mansour$^1$}
\email{andrew.mansour@student.unsw.edu.au}
\affiliation{$^1$School of Physics, University of New South Wales,
Sydney 2052, Australia}
\affiliation{$^2$Helmholtz Institute Mainz, Johannes Gutenberg University, 55099 Mainz, Germany}

\begin{abstract}
Nuclei with an octupole deformation have a non-zero electric octupole moment, electric dipole moment (EDM), Schiff moment and magnetic quadrupole moment (MQM) in the intrinsic frame which rotates with the nucleus. In a state with definite angular momentum in the laboratory frame, these moments are forbidden by parity (P) and time reversal invariance (T) conservation, meaning their expectation values vanish due to nuclear rotation. However, nuclei with an octupole deformation have doublets of close opposite parity rotational states with the same spin, which are mixed by T,P-odd nuclear forces. This mixing produces the orientation of the nuclear axis along nuclear spin and all moments existing in the intrinsic frame appear in the laboratory frame (provided the nuclear spin $I$ is sufficiently large  to allow such moment). Such a mechanism produces enhanced T,P-violating nuclear moments. This enhancement also takes place in nuclei with a soft octupole vibration mode. Schiff moments in such nuclei have been calculated in previous works. In the present paper we consider the magnetic quadrupole moment which appears in isotopes with nuclear spin $I \ge 1$. Magnetic interaction between the nuclear MQM and electrons produces an atomic EDM and T,P-violating nuclear spin - molecular axis interaction constants for molecules in electronic states with non-zero electron angular momentum.  Measurements of these constants may be used to test CP-violation theories and search for axion dark matter in atomic, molecular and solid state experiments. Potential candidate nuclei include $^{153}$Eu, $^{161}$Dy, $^{221}$Fr, $^{223}$Fr, $^{223}$Ra, $^{223}$Rn, $^{225}$Ac, $^{227}$Ac, $^{229}$Th, $^{229}$Pa, $^{233}$U and $^{235}$U. We subsequently consider molecules containing these nuclei (EuO, EuN$^+$, RaF, AcO, AcN$^{+}$, AcF$^{+}$, ThO and ThF$^+$).

\end{abstract}
\date{\today}
\maketitle

\section{Introduction}

\subsection{Nuclear magnetic quadrupole moments produced by T,P-odd nuclear forces and atomic and molecular EDM produced by magnetic  field of these  moments}

Measurements  of  atomic and molecular time reversal (T) and parity (P) violating electric dipole moments are used to test unification theories predicting CP-violation. Such measurements have already excluded a number of models and significantly reduced the parametric space of other popular models including supersymmetry~\cite{PR,ERK}. Another motivation is related to the baryogenesis problem, the matter-antimatter asymmetry in the universe which is produced by an unknown CP-violating interaction. The expected magnitude of an  EDM is  very small, therefore, it is advantageous to search for mechanisms that enhance the effects - see e.g.~\cite{Khriplovich,KL,GF}. 

Schiff demonstrated that the nuclear EDM is completely screened  in neutral atoms and molecules, and noted that a nonzero atomic EDM still may be produced if the distribution of EDM and charge in a nucleus are not proportional to each other~\cite{Schiff}.  Further works~\cite{Sandars,Hinds,SFK,FKS1985,FKS1986} introduced and calculated the so called Schiff moment, a vector moment producing an electric field inside the nucleus after taking into account the screening of the nuclear EDM by electrons. The electric field produced by this Schiff moment polarizes the atom and produces an atomic EDM directed along the axis of nuclear spin. Refs.~\cite{Sandars,Hinds} calculated the Schiff moment due to the proton EDM. Refs.~\cite{SFK,FKS1985,FKS1986} calculated (and named) the nuclear Schiff moment produced by the P,T-odd nuclear forces. 
It was shown in~\cite{SFK} that the contribution of the T,P-odd forces to the nuclear EDM and Schiff moment is larger than the contribution of a nucleon EDM. In Ref.~\cite{FG} an accurate expression for the Schiff moment electrostatic potential has been derived and the finite nuclear size corrections to the Schiff moment operator introduced (see also \cite{AKozlovC,AKozlovA}). The Schiff moment is proportional to the third power of the nuclear size which is very small on an atomic scale.

The magnetic interaction between nuclear MQM and atomic electrons mixes electron orbitals of opposite parity and produces an atomic EDM and T,P-violating nuclear spin - molecular axis interaction constants for molecules in electronic states with non-zero electron angular momentum \cite{SFK}. This magnetic interaction is not screened, so generically (without special enhancement factors) the atomic EDM produced by the interaction between nuclear MQM and electrons is expected to be an order of magnitude bigger than the EDM produced by the Schiff moment \cite{SFK} and two orders of magnitude bigger than the EDM produced by the electric octupole moment \cite{Octupole}. The nuclear EDM, Schiff, electric octupole and magnetic quadrupole produced by the T,P-odd nuclear forces are enhanced due to an opposite parity level with the same spin close to the ground state \cite{HH,SFK,Octupole}. Collective enhancement of the magnetic quadrupole moments in nuclei with a quadrupole deformation has been demonstrated in \cite{F1994} (see also \cite{FDK,Lackenby2018}).

\subsection{Enhancement of nuclear magnetic quadrupole moment due to nuclear octupole deformation and soft octupole vibration mode}

The largest enhancement ($\sim 10^2 - 10^3$ times) of the nuclear Schiff moment and electric octupole moment occurs in nuclei with an intrinsic octupole deformation, where both the small energy difference of nuclear levels with opposite parity and the collective effect work together \cite{Auerbach,Spevak,Octupole}. According to Refs.~\cite{Auerbach,Spevak} this happens in some isotopes of Fr, Rn, Ra and other actinide atoms. The atomic and molecular EDMs produced by the Schiff moment, electric octupole moment and MQM increase with nuclear charge $Z$ at a faster rate than $Z^2$  \cite{SFK,Octupole}. This again explains why the EDMs in actinide atoms and their molecules are expected to be significantly larger than in other systems.


The T,P-violating electric octupole moment and Schiff moment in the laboratory frame are proportional to the squared octupole deformation parameter $(\beta_3)^2$ \cite{Spevak,Octupole}. In nuclei with an octupole deformation  $(\beta_3)^2 \sim (0.1)^2$. According to Ref. \cite{Engel2000}, in nuclei with a soft octupole vibration mode the squared dynamical octupole deformation parameter $\ev{(\beta_3)^2} \sim (0.1)^2$, i.e. it is the same as the static octupole deformation. This implies that a similar enhancement of the Schiff moment and electric octupole moment may be due to the dynamical octupole effect \cite{Engel2000,FZ,soft2} in nuclei where $\ev{\beta_3}=0$ \footnote{Recall an ordinary oscillator where $\ev{x}=0$ while $\ev{x^2} \neq 0$ .}. This observation significantly increases the list of nuclei where the Schiff moment and electric octupole moment are enhanced.

In the papers \cite{Auerbach,Spevak,EngelRa,Jacek2018}, numerical calculations of the Schiff moments and estimates of the atomic EDM produced by the electrostatic interaction between electrons and these moments have been done  for  $^{223}$Ra, $^{225}$Ra, $^{223}$Rn, $^{221}$Fr, $^{223}$Fr, $^{225}$Ac and $^{229}$Pa. Unfortunately, these nuclei have a short lifetime. Several experimental groups have considered experiments with $^{225}$Ra and $^{223}$Rn \cite{RaEDM,RaEDM2,RnEDM}. The only published EDM measurements  \cite{RaEDM,RaEDM2} have been done for $^{225}$Ra, which has a half-life of 15 days. In spite of the Schiff moment enhancement in $^{225}$Ra, EDM measurements have not yet reached the sensitivity of the T,P-odd interaction comparable to the Hg EDM experiment \cite{HgEDM}. These experiments continue, despite the problems caused by the instability of $^{225}$Ra and the relatively small number of atoms available. In Ref. \cite{Th} the    
nuclear Schiff moment of $^{229}$Th has been estimated, as this nucleus has a much longer lifetime (7917 years). Ref. \cite{FF19} extended the list of the candidates for the enhanced Schiff moments to include the following
stable isotopes: $^{153}$Eu, $^{161}$Dy, $^{163}$Dy, $^{155}$Gd, and long lifetime nuclei   $^{235}$U, $^{237}$Np, $^{233}$U, $^{229}$Th, $^{153}$Sm, $^{165}$Er, $^{225}$Ac, $^{227}$Ac, $^{231}$Pa, $^{239}$Pu.

In this paper we estimate the nuclear MQM in nuclei where we expect $\ev{(\beta_3)^2}\,  > (0.05)^2$ and express the MQM in terms of the CP-violating $\pi$-meson - nucleon interaction constants ${\bar g}_0$, ${\bar g}_1$, ${\bar g}_2$, QCD parameter ${\bar \theta}$ and quark chromo-EDMs. 
Potential candidate nuclei include $^{153}$Eu $(I^{P} = 5/2^{+})$,  $^{161}$Dy $(5/2^{+})$, $^{221}$Fr $(5/2^{-})$, $^{223}$Fr $(3/2^{-})$, $^{223}$Ra $(3/2^{+})$ , $^{223}$Rn $(7/2^{+})$, $^{225}$Ac $(3/2^{-})$,  $^{227}$Ac $(3/2^{-})$,  $^{229}$Th $(5/2^{+})$, $^{229}$Pa $(5/2^{+})$,  $^{233}$U $(5/2^{+})$,  $^{235}$U $(7/2^{-})$ and $^{237}$Np $(5/2^{+})$. We then estimate the T,P-odd effects in molecules containing these nuclei (EuO, EuN$^+$, RaF, AcO, AcN$^{+}$, AcF$^{+}$, ThO and ThF$^+$).



\subsection{Oscillating nuclear magnetic quadrupole moments and atomic and molecular electric dipole moments produced by axion dark matter}

The CP-violating neutron EDM may be due to the QCD $\theta$-term \cite{Witten}.  It was noted in Ref.~\cite{Graham} that axion dark matter produces an oscillating neutron EDM, as the axion field is equivalent to the oscillating ${\bar \theta}$. The QCD $\theta$-term also produces T,P-odd nuclear forces, creating T,P-odd nuclear moments. Correspondingly, the axion field  also produces oscillating nuclear T,P-odd moments \cite{Stadnik} which are  enhanced  by the octupole mechanism. To obtain results for the oscillating T,P-odd moments it is sufficient to replace the constant ${\bar \theta}$ by ${\bar \theta}(t)= a(t)/f_a$, where $f_a$ is the axion decay constant, $a(t) =a_0 \cos{m_a t}$, $(a_0)^2= 2 \rho /(m_a)^2$, where $\rho$ is the axion dark matter energy  density \cite{Graham,Stadnik}.  Moreover, in the case of a resonance between the frequency of the axion field oscillations and molecular transition frequency there may be an enormous resonance enhancement of the oscillating nuclear T,P-odd  moment  effect  \cite{OscillatingEDM}.  Since oscillating nuclear T,P-odd moments may be produced by axion dark matter, corresponding measurements may be used to search for this dark matter. This research is in progress, and the first results have been published in Ref. \cite{nEDM}, in which the oscillating neutron EDM and oscillating $^{199}$Hg Schiff moment have been measured.  The effect produced by the oscillating axion-induced Pb Schiff moment in solid state materials has been measured by the CASPEr collaboration in Ref. \cite{CasperNew}.
The effect of oscillating T,P-odd nuclear polarizability has been measured in Ref. \cite{Cornell} (see theory in \cite{pol1,pol2,pol3}). Oscillating MQMs produce resonance transitions in molecules \cite{MQMtransitions}.

\section{Estimates of  nuclear magnetic quadrupole moments} \label{Schiff}
\subsection{Calculation of MQM  in nuclei with octupole deformation or soft octupole mode}
The magnetic quadrupole moment of a nucleus due to the electromagnetic current of a single nucleon with mass $m$ is defined by the second order tensor operator \cite{SFK},
\begin{align} \label{eq:MQMTensor}
\begin{split}
\hat{M}_{kn}^{\nu} = \dfrac{e}{2m}\biggl[3\mu_{\nu}\left(r_k\sigma_n + \sigma_kr_n - \dfrac{2}{3}\delta_{kn}\hat{\boldsymbol{\sigma}}\textbf{r}\right)  \\
 + 2q_{\nu}\left(r_kl_n + l_kr_n\right)\biggr]
\end{split}
\end{align}
where $\nu = p,n$ for protons and neutrons respectively and m, $\mu_{\nu}$ and $q_{\nu}$ are the mass, magnetic moment in nuclear magnetons $\dfrac{e}{2m}$ and charge of the nucleon respectively.  In the case of an axially symmetric nucleus for an orbital with definite spin projection on the nuclear axis $\Sigma$ and orbital angular momentum projection $\Lambda$, we may give the following estimate for the expectation value of the MQM
\begin{align} \label{eq:Mzz}
M_{zz}^{\nu} = 4 \dfrac{e}{2m}\mu_z  \ev{r_z}, 
\end{align}
where $\mu_z= 2 \Sigma \mu_{\nu} + q_{\nu} \Lambda$ is the projection of the nucleon magnetic moment  and $\ev{ r_z}$ is the expectation value of the radius vector ${\bf r}$. The latter vanishes in the absence of an octupole deformation, therefore $\ev{ r_z} \sim \beta_3 R$, where $R\approx A^{1/3} 1.2$ fm is the nuclear radius.  

Above, we have presented the MQM in the intrinsic frame which rotates with the nucleus. We must now find the MQM in the laboratory frame. A nucleus with an octupole deformation and non-zero nucleon angular momentum has a doublet of close opposite parity rotational states $\ket{I^{\pm}}$ with the same angular momentum $I$ 
\begin{align} \label{doublet}
\ket{ I^{\pm} }=\frac{1}{\sqrt{2}} (\ket{\Omega} \pm \ket{-\Omega}), 
\end{align}
where $\Omega=\Sigma +\Lambda$ is  the projection of $I$ on to the nuclear axis). In the case of the ordinary electric quadrupole moment $Q$, which conserves $T$ and $P$ symmetries, we have the relation $\mel{\Omega }{ Q_{zz} }{ \Omega}= \mel{ - \Omega }{ Q_{zz} }{ - \Omega}$, and the following relation between the intrinsic value  $Q_{zz}$ and laboratory value $Q$ in the ground rotational state \cite{BohrMott}:
 \begin{align} \label{eq:RotationalFactor}
Q = \dfrac{I\left(2I - 1\right)}{\left(I + 1 \right)\left(2I + 3\right)}Q_{zz},
\end{align}
where $I=I_z= \left|\Omega\right|$ is the projection of total nuclear angular momentum (nuclear spin $I$) on the symmetry axis, $\Omega={\bf I \cdot n}$. This expression for $Q$ shows that we can only detect these second order tensor properties in nuclei with spin $I > 1/2$. In the case of the MQM we have $\mel{\Omega }{ M_{zz} }{ \Omega}= - \mel{ - \Omega }{ M_{zz} }{ - \Omega}$, and  the laboratory value of $M$ vanishes in the states of definite parity (\ref{doublet}) which have equal weights of the $\Omega$ and   $-\Omega$ components. This is a consequence  of $T$ and $P$ conservation. 

However, the states of  this doublet are mixed by the $T,P-$violating interaction $W$, with mixing coefficient:
\begin{equation}\label{alpha}
 \alpha_{\pm}=\frac{\mel{I^-}{ W}{ I^+}}{E_+  -  E_-} . 
\end{equation}
This mixing produces non-equal weights of $\Omega$  and  $-\Omega$ ,  $(1+  \alpha_{\pm})^2/2$ and $(1 - \alpha_{\pm})^2/2$ respectively, and leads to a non-zero expectation value of $\ev{{\bf I \cdot n}}$, i.e.  the mixing  polarises the  nuclear axis ${\bf n}$ along the nuclear spin ${\bf I}$ \cite{Auerbach,Spevak}:
\begin{equation}\label{n}
 \ket{n_z}= 2 \alpha_{\pm} \frac{I_z}{I+1}\,.
\end{equation}
As a result  all intrinsic  T,P-odd nuclear moments appear in the laboratory frame. Using Eqs. (\ref{eq:Mzz}--\ref{eq:RotationalFactor}), we obtain the following for MQMs in the ground nuclear state
\begin{align} \label{eq:MQMRotationalFactor}
\begin{split}
M & = 2 \alpha_{\pm} \dfrac{I\left(2I - 1\right)}{\left(I + 1 \right)\left(2I + 3\right)}  M_{zz} \ , \\
& = 8 \alpha_{\pm} \dfrac{I\left(2I - 1\right)}{\left(I + 1 \right)\left(2I + 3\right)}   \dfrac{e}{2m}\mu_z \ev{ r_z}\,.
\end{split}
\end{align}
  According to Ref. \cite{Spevak} the T,P-violating matrix element is approximately equal to
   \begin{equation}\label{W}
   \mel{I^-}{ W}{ I^+} \approx \frac{\beta_3 \eta}{A^{1/3}} [ \textrm{eV}].
  \end{equation}  
  Here $\eta$ is the dimensionless strength constant of the nuclear $T,P$- violating potential $W$:
   \begin{equation}\label{eta}
 W= \frac{G}{\sqrt{2}} \frac{\eta}{2m} ({\bf \sigma \nabla}) \rho ,
   \end{equation}
where $G$ is the Fermi constant, $m$ is the  nucleon mass and $\rho$ is the nuclear number density. The nuclear magnetic moment in the intrinsic frame is related to $\mu$ in the laboratory frame by
\begin{equation}\label{mu}
 \mu=  \frac{I}{I+1}\mu_z\,.
\end{equation}
The value of $\mu$ has been measured in all nuclei of experimental interest. Using   $\ev{r_z} \approx  1.2 \beta_3 A^{1/3} $ fm we obtain  nuclear MQM 
\begin{align} \label{eq:Mlab}
M\approx  \dfrac{2I - 1}{2I + 3} \ev{(\beta_3)^2}  \frac{  \textrm{eV}} {E_+  -  E_-}  \mu \eta \, e \,\textrm{fm}^2\,.
\end{align}
The typical energy interval of the doublet  $E_+  -  E_-$ is between 25 keV and  100 keV, which is much smaller than the interval between the opposite parity orbitals in spherical nuclei ($\sim$ 8 MeV). Therefore, value of the MQM in nuclei with an octupole deformation may be  1 - 2 orders of magnitude bigger than the MQM of a spherical nucleus estimated in Ref. \cite{SFK}. To avoid misunderstanding, we should mention that this enhancement is not as big as the enhancement of the nuclear Schiff moment (2-3 orders of magnitude)  which has a collective nature in the nuclei with an octupole deformation  \cite{Auerbach,Spevak,EngelRa,Jacek2018}.

Note that the MQM in Eq. (\ref{eq:Mlab})  is quadratic in the octupole deformation parameter. Therefore, it is sufficient to have a soft octupole deformation mode, i.e. dynamical deformation, which actually gives values of $ \ev{(\beta_3)^2} $ comparable to that for the static octupole deformation. In this case, the situation is similar to the Schiff moment calculations, see Refs. \cite{Engel2000,FZ}.

Within the meson exchange theory, the $\pi$-meson exchange gives the dominating contribution to the T,P-violating nuclear forces \cite{SFK}. According to Ref. \cite{FDK} the neutron and proton constants in the T,P-odd potential (\ref{eta}) may be presented as  
\begin{align}
\eta_n=-\eta_p =   (-  g \bar{g}_{0} + 5 g \bar{g}_{1} + 2 g  \bar{g}_{2} ) 10^{6}, 
\end{align} 
where $g$ is the strong $\pi$-meson - nucleon interaction constant and ${\bar g}_0$, ${\bar g}_1$, ${\bar g}_2$ are the $\pi$-meson - nucleon CP-violating interaction constants in the isotopic channels $T=0,1,2$. The numerical coefficient comes from $(Gm_{\pi}^2/2^{1/2})^{-1}= 6.7 \cdot 10^6$ times the factor 0.7 corresponding to the zero range reduction of the finite range interaction due to the $\pi_0$ exchange \cite{SFK,FKS1985,FKS1986}.


We can also express $\eta$ in terms of the QCD $\theta$-term constant. Using the results presented in Refs. ~\cite{Yamanaka2017,Vries2015,Bsaisou2015}
\begin{align}
g \bar{g}_{0} &=-0.21  \, \bar{\theta}, \\
g \bar{g}_{1} &= 0.046  \,  \bar{\theta}, 
\end{align}
 we obtain  
\begin{align} \label{etawrttheta}
 \eta_n=-\eta_p =  4 \times 10^{5} \ \bar{\theta},
\end{align}
Further, we can express $\eta$ via the quark chromo-EDMs ${\tilde d_u}$ and  ${\tilde d_d}$. Using relations  $g {\bar g}_0 = 0.8 \times 10^{15}({\tilde d_u} +{\tilde d_d})$/cm,  $g {\bar g}_1 = 4 \times 10^{15}({\tilde d_u} - {\tilde d_d})$/cm~\cite{PR} we obtain:
\begin{align}
 \eta_n=-\eta_p =  ( 2 (\tilde{d_{u}} - \tilde{d_{d}}) - 0.1 (\tilde{d_{u}} + \tilde{d_{d}}) ) 10^{22}/\text{cm}.
\end{align}
In the expressions above, the interaction constants have opposite signs for a valence proton or valence neutron (which in this case is an unparied nucleon, defining a non-zero value of $\Omega$). However, the magnetic moments of protons and neutrons have an opposing sign, meaning the overall sign of the product  $\mu \eta$  for valence protons and neutrons is the same. As we are interested in  the limits on these interaction constants obtained from the limits on the measured values of MQM measurements, which are presented as absolute values, this sign of these constants is not of importance. Therefore, in our estimates of MQM for specific nuclei we are only interested in their absolute values. Using Equation (\ref{eq:Mlab}) in conjunction with the above calculations, we may now write the MQM in terms of the interaction constants

\begin{align} 
M(\eta) & \approx 10^{-31}M_0 \eta\,,
\label{Meta}  \\
M(\theta) & \approx 4  \cdot 10^{-26} M_0 \bar{\theta}\,,
\label{Mtheta} \\
M(g) & \approx 10^{-25} M_0 \,
\label{Mg} (-  g \bar{g}_{0} + 5 g \bar{g}_{1} + 2 g  \bar{g}_{2} ) \,, \\
M(d) & \approx 10^{-9} M_0 \,
\label{Md} ( 2  (\tilde{d_{u}} - \tilde{d_{d}})-0.1 (\tilde{d_{u}} + \tilde{d_{d}}) )/\textrm{ cm} \,,
\end{align}
where
\begin{align}
M_0  & = \dfrac{2I - 1}{2I + 3} (\beta_3)^2  \frac{  100 \, \textrm{keV}} {E_+  -  E_-} \mu \, e \,  \textrm{ cm}^2.  
\end{align}


Note that a comparable contribution to the MQM is given by a spin hedgehog mechanism which requires quadrupole deformation (but does not require octupole deformation) \cite{F1994}.
Magnetic quadrupoles in this case have collective nature, somewhat similar to collective electric quadrupoles in deformed nuclei.  Collective contributions to MQM have been calculated in  Refs. \cite{F1994,FDK,Lackenby2018}.

\subsection{MQM for specific nuclei}

We may now present our calculations of the MQM for specific nuclei with an octupole deformation or soft octupole vibration mode. As an example, let us first consider the $^{237}$Np nucleus, which has a half-life of 2.14 million years, and is produced in macroscopic quantities in nuclear reactors. The interval between opposite parity levels which are mixed by the T,P-odd interaction is $E(5/2^{-}) - E(5/2^{+}) = 59.5$ keV. The experimental nuclear excitation spectra for this nucleus satisfies the criteria for the existence of an octupole deformation. Thus, given $^{237}$Np has one additional proton above $^{236}$U, it's octupole deformation parameter may be interpolated between the values for $^{234}$U and $^{238}$U (see Ref.~\cite{Afanasjev2016}) to be $\beta_{3} = 0.12$. Taking the value of the nuclear magnetic moment from Ref.~\cite{StoneTable} to be $\mu = 3.1(4)$, we may calculate explicitly the MQM for $^{237}$Np via substitution into Equations (\ref{Meta} - \ref{Md}) above

\begin{align} \label{MQMNp}
\begin{split}
M(\eta) & \approx 3.8 \cdot 10^{-33} \eta \, e \, \textrm{ cm}^2,  \\
M(\theta) & \approx 1.5  \cdot 10^{-27}  \bar{\theta} \, e\, \textrm{ cm}^2,  \\
M(g) & \approx 3.8 \cdot 10^{-27}   
(-  g \bar{g}_{0} + 5 g \bar{g}_{1} + 2 g  \bar{g}_{2} ) \, e \,  \textrm{ cm}^2,  \\
M(d) & \approx  3.8 \cdot 10^{-11} ( 2  (\tilde{d_{u}} - \tilde{d_{d}})-0.1 (\tilde{d_{u}} + \tilde{d_{d}}) ) \, e \,  \textrm{ cm}. 
\end{split}
\end{align}

Using a similar method, we performed these calculations for a range of nuclei which exhibit a nuclear excitation spectra consistent with that of an octupole deformation: $^{153}$Eu, $^{161}$Dy, $^{221}$Fr, $^{223}$Fr, $^{223}$Ra, $^{223}$Rn, $^{225}$Ac, $^{227}$Ac, $^{229}$Th, $^{229}$Pa, $^{233}$U and $^{235}$U. The results are presented in Table \ref{Nuclei1}. Values for the nuclear magnetic moments are taken from Ref.~\cite{StoneTable}, unless otherwise indicated.

\begin{widetext}
\begin{center}
\begin{table}[ht]
\begin{ruledtabular}
\begin{tabular}{ccccccccc} 
& & & & & \multicolumn{4}{c}{$M$}  \\
         \cmidrule{6-9} 
Nucleus & $I^{P}$ & $\mu$ (nm) & $\Delta E$ (keV) & $\beta_{3}$ & $  10^{-34} \ \eta \ e  \ \text{cm}^{2}$ & $   10^{-28} \bar{\theta} \ e \ \text{cm}^{2} $ &  $   10^{-28} \  \bar{G} \ e  \ \text{cm}^{2}$~\footnote{For brevity, we make the substitution $\bar{G} \equiv (- g \bar{g_{0}} + 5 g \bar{g_{1}} + 2 g \bar{g_{2}})$. } & $   10^{-12} \  \tilde{D} \ e \ \text{cm} $~\footnote{Similarly, we denote $\tilde{D} \equiv 2(\tilde{d}_{u} - \tilde{d}_{d}) - 0.1 (\tilde{d}_{u} + \tilde{d}_{d}))$.}   \\
\midrule
$^{153}$Eu & 5/2$^{+}$ & 1.5(3) & 97.4  & 0.15~\cite{FF19} & 18 & 7.1 & 18 & 18    \\
$^{161}$Dy & 5/2$^{+}$ & -0.48(3) & 25.7  & $\sim 0.1$ & -9.4 & -3.7 & -9.4 & -9.4    \\
$^{221}$Fr & 5/2$^{-}$ & 1.6(3) & 234  & 0.10~\cite{Spevak} & 3.4 & 1.4 & 3.4 & 3.4    \\
$^{223}$Fr & 3/2$^{-}$ & 1.2(2) & 160  & 0.090~\cite{Spevak} & 2.0 & 0.79 & 2.0 & 2.0    \\
$^{223}$Ra & 3/2$^{+}$ & 0.27(2) & 50.1  & 0.10~\cite{Spevak} & 1.8 & 0.72 & 1.8 & 1.8    \\
$^{223}$Rn & 7/2$^{+}$ & -0.78(8) & 130  & 0.081~\cite{Spevak} & -2.3 & -0.94 & -2.3 & -2.3    \\
$^{225}$Ac & 3/2$^{-}$ & 1.1(1)~\footnote{Using the value for $^{227}$Ac, which has identical spin and parity to $^{225}$Ac.}  &  40.1  & 0.10~\cite{Spevak} & 9.9 & 4.0 & 9.9 & 9.9    \\
$^{227}$Ac & 3/2$^{-}$ & 1.1(1) & 23.4  & 0.12~\cite{FF19} & 13 & 5.0 & 13 & 13    \\
$^{229}$Th & 5/2$^{+}$ & 0.46(4) & 133  & 0.12~\cite{Minkov} & 2.3 & 0.91 & 2.3 & 2.3    \\
$^{229}$Pa & 5/2$^{+}$ & 1.96~\footnote{Calculated using the Schmidt model.} & 0.220  & 0.082~\cite{Spevak} &  3000& 1200 & 3000  & 3000    \\
$^{233}$U & 5/2$^{+}$ & 0.59(5) & 299  & 0.17~\cite{Afanasjev2016} & 4.8 & 1.9 & 4.8 & 4.8    \\
$^{235}$U & 7/2$^{-}$ & -0.38(3) & 81.7  & 0.17~\cite{FF19} & -8.1 & -3.2 & -8.1 & -8.1   \\
$^{237}$Np & 5/2$^{+}$ & 3.1(4) & 59.5  & 0.12~\cite{Afanasjev2016} & 38 & 15 & 38 & 38   \\
\end{tabular}
\end{ruledtabular}
\caption{Nuclear spin $I$, magnetic moment $\mu$, experimental interval between opposite parity levels mixed by the T,P-odd interaction $\Delta E = E^{+} - E^{-}$ and octupole deformation parameter $\beta_{3}$. Calculations using these quantities allow for the MQM of various octupole deformed nuclei to be expressed in terms of different fundamental constants.}
\label{Nuclei1}
\end{table}
\end{center}
\end{widetext}

\section{MQM energy shift in diatomic molecules} \label{sec:MQMmolecule}

The direct measurement of nuclear MQMs in an external magnetic field is unfeasible. As mentioned above the use of  molecular systems is promising, as the nuclear MQM will interact with the internal magnetic field. Molecules in particular present the most attractive option due to existence of very close paired levels of opposite parity, the  $\Omega$-doublet - see e.g.  \cite{SFK}. For polar molecules consisting of a heavy and light nucleus (for example, Th and O) the effect of MQM is $\sim Z^2$,  and thus it is calculated for the heavier nucleus. The Hamiltonian of diatomic paramagnetic molecule including the $T, P-$ odd nuclear moment effects is given by \cite{SFK,Kozlov1995}:

\begin{align}
H = W_d d_e \mathbf{S}\cdot\mathbf{n} + W_{Q}\dfrac{Q_s}{I}\mathbf{I}\cdot\mathbf{n} - \dfrac{W_{M}M}{2I(2I -1)}\mathbf{S}\hat{\mathbf{T}}\mathbf{n},
\end{align}
where $d_e$ is the electron EDM, $Q_s$ is the nuclear Schiff moment, $M$ is the nuclear MQM, $\mathbf{S}$ is the effective  electron spin, $\mathbf{n}$ is the symmetry axis of the molecule, $\hat{\mathbf{T}} $ is the second rank tensor operator characterised by the nuclear spins $T_{ij} = I_iI_j + I_jI_i - \tfrac{2}{3}\delta_{ij}I(I + 1)$  and  $W_d$, $W_Q$ and $W_M$ are fundamental parameters for each interaction which are dependent on the particular molecule. We have omitted the $T$-,$P$- odd electron-nucleon interaction terms which are presented e.g. in the review \cite{GF}.

Parameters $W_d$, $W_Q$ and $W_M$
are related to the electronic molecular structure of the state.
These parameters are calculated using relativistic many-body methods (see e.g. \cite{FDK}).  For the nuclear MQM we are interested only in $W_{M}$, which has been calculated for various molecules of our interest: RaF~\cite{Talukdar}, AcO~\cite{Oleynichenko2022}, AcN$^{+}$~\cite{Oleynichenko2022}, AcF~\cite{Oleynichenko2022}, ThO \cite{Skripnikov2014} and ThF$^{+}$\cite{Skripnikov2017}. For other molecules of interest, where accurate many-body calculations are absent (such as EuO and EuN$^{+}$), we may use the $Z^2$ scaling for the electron factor in the energy shifts $W_{M} \Omega$, and thus estimate the value of $W_{M}$. We may thus express the energy shifts induced by MQMs in terms of the $CP$-violating $\pi$-meson - nucleon interaction constants $\bar{g}_{0}$, $\bar{g}_{1}$ and $\bar{g}_{2}$, the QCD parameter $\theta$ and the quark chromo-EDMs. The results of these calculations are shown in Table \ref{MoleculeTable}.



Using the current limits on the $CP$-violating parameters $|\bar{\theta}| < 2.4 \times 10^{-10}$,  $|\tilde{d}_{u} - \tilde{d}_{d} | < 6 \times 10^{-27}$cm and  $|\frac{1}{2}\tilde{d}_{u} + \tilde{d}_{d} | < 3 \times 10^{-26}$cm \cite{Swallows}, the MQM energy shifts $(|W_{M} M S|)$ in $^{229}$ThO are $< 25 \ \mu$Hz and $< 23 \ \mu$Hz respectively. The $^{232}$ThO molecule has recently been used
to set new limits on the electron EDM, with an accuracy for the energy shift of $80 \ \mu$Hz \cite{ACMECollaboration}, which is a factor of 12 improvement in accuracy compared to their previous work \cite{TheEDM}.

$^{232}$Th has zero nuclear spin and no MQM. The possibility of performing a similar experiment with $^{229}$ThO is promising, as in principle, such measurements would improve constraints on nuclear $CP$-violating interactions. The results for the other molecules presented in Table \ref{MoleculeTable} are similar, and given the expected enhancement of the MQM in their nuclei, these may be useful candidates in the search for new physics in the hadronic sector. 
\begin{widetext}
\begin{center}
\begin{table}[!ht]
\begin{ruledtabular}
    \begin{tabular}{cccccccc}
         & & & $| W_{M} | $ & \multicolumn{4}{c}{$|W_{M} MS| $ ($\mu$Hz)}  \\
         \cmidrule{4-8} 
         Molecule & $I^{P}$ & State & $10^{39} \ \mu \text{Hz}/ e \cdot \text{cm}^{2}$ & $10^{5} \ \eta$ & $10^{11} \ \bar{\theta}$ & $10^{11} \bar{G}$ & $10^{27} \ \tilde{D} / \text{cm}$ \\
         \midrule
         $^{153}$EuO,$^{153}$EuN$^{+}$  & $5/2^{+}$ & --~\footnote{At this time, the $\Omega$ value of these Eu based molecules are unknown.} & 0.60~\footnote{Estimated using $Z^{2}$ scaling.} & 11 & 4.2 & 11 & 11 \\ 
         $^{223}$RaF & $3/2^{+}$ & $^{2}\Sigma_{1/2}$ & 1.2~\cite{Talukdar} & 1.1 & 0.44 & 1.1 & 1.1 \\
         $^{225}$AcO, $^{225}$AcN$^{+}$, $^{225}$AcF$^{+}$ & $3/2^{-}$ & $^{2}\Sigma_{1/2}$ & 1.7~\cite{Oleynichenko2022} & 8.2 & 3.3 & 8.2 & 8.2 \\
         $^{227}$AcO, $^{227}$AcN$^{+}$, $^{227}$AcF$^{+}$ & $3/2^{-}$ & $^{2}\Sigma_{1/2}$ & 1.7~\cite{Oleynichenko2022} & 11 & 4.4 & 11 & 11 \\
         $^{229}$ThO & $5/2^{+}$ & $^{3}\Delta_{1}$ & 1.1~\cite{Skripnikov2014} & 2.5 & 1.0 & 2.5 & 2.5 \\
         $^{229}$ThF$^{+}$ & $5/2^{+}$ & $^{3}\Delta_{1}$ & 0.59~\cite{Skripnikov2017} & 1.3 & 0.53 & 1.3 & 1.3 \\
    \end{tabular}    
    \end{ruledtabular}
    \caption{Frequency shifts due to the MQM interaction with the electron magnetic field of the molecules. The energy shifts are presented in terms of the CP-violating parameters $\eta$, $\bar{\theta}$, $\bar{G} \equiv (- g \bar{g_{0}} + 5 g \bar{g_{1}} + 2 g \bar{g_{2}})$ and  $\tilde{D} \equiv 2(\tilde{d}_{u} - \tilde{d}_{d}) - 0.1 (\tilde{d}_{u} + \tilde{d}_{d})$.}
    \label{MoleculeTable}
\end{table}
\end{center}
\end{widetext}

\section{Conclusion}
In this publication we have presented a method to estimate the magnetic quadrupole moment in various isotopes with nuclear spin $I \ge 1$ and an experimental nuclear excitation spectra indicating an octupole deformation. Such nuclei exhibit an enhanced MQM due to the mixing of close, opposite parity rotational states with identical spin, and their study may provide an opportunity for  detection of T,P-odd effects in the hadronic sector. Paramagnetic molecules containing heavy atoms are  promising candidates for measuring the nuclear MQM. For several molecules of our interest, there exist accurate many-body calculations of the electronic factors $W_{M}$ before the nuclear magnetic quadrupole moment. Using these values, we calculated the energy shift induced by MQM in these molecules. In molecules where these calculations are absent, we have used the scaling $W_{M} \propto Z^{2}$ and accurate results for other molecules to estimate  $W_{M}$.  The calculations presented in this work may allow experimentalists to measure the values of fundamental $CP$-violating $\pi$-meson - nucleon interaction constants $\bar{g}_{0}$, $\bar{g}_{1}$ and $\bar{g}_{2}$, the QCD parameter $\theta$ and the quark chromo-EDMs.

We should note that for planning and interpretation of future experiments one needs dedicated many-body calculations of the nuclear MQM and their effects in molecules. Molecular calculations have already achieved few  percent accuracy in diatomic molecules \cite{Oleynichenko2022,Skripnikov2014,Skripnikov2017}. Recently high accuracy molecular calculations of MQM effects  have been extended to triatomic molecules \cite{Maison2020,Maison2019}. There are no 
accurate nuclear many body calculations of MQM. However, the methods of the calculations of the nuclear Schiff moments presented in Refs.\cite{EngelRa,Jacek2018} are applicable for the calculations of MQM. Moreover, we may expect that calculations of the nuclear MQM may be more accurate since in the formula for the Schiff moment there are two terms of opposite sign while for MQM we do not expect any cancellations. 

\section{Acknowledgements}
This work was supported  by the Australian Research Council grants DP190100974 and  DP200100150, and the JGU Gutenberg Fellowship.

\bibliographystyle{apsrev}

\end{document}